\documentclass[11pt]{cernrep}
\usepackage{graphicx}
\usepackage{cite,./mcite}

\usepackage{amsmath,amssymb,bm}
\begin{document}
\title{
\vspace*{-2cm}
\begin{flushright}
{\rm \small
OUTP-08-22-P}\\
\end{flushright}
\vspace*{2cm}
Nonperturbative corrections from  an s-channel  approach } 
\author{F. ~Hautmann}
\institute{Department  of  Theoretical Physics, University of Oxford,  
  Oxford OX1 3NP }
\maketitle
\begin{abstract}
We   report on studies  
 of multi-parton corrections from  nonlocal operator expansion. 
 We    discuss  relations between  eikonal-line matrix elements and 
 parton distributions, and present an illustration for initial-state collinear 
 evolution.  
\end{abstract}

\hskip 1.0  cm  
{\em Contribution to   ``HERA and the LHC'' 
  Workshop Proceedings, CERN, 
 2008 }

\section{Introduction}
\label{sec:intro}

Nonperturbative dynamics affects the structure of 
LHC  events  even  for   high momentum transfer,   
 through      hadronization, soft  underlying scattering,  
    multiple hard interactions.  Models for 
 these   processes   are   necessary, for instance,  
 for  Monte Carlo   generators  to  produce 
 realistic     event   simulations.

The treatment  of multiple  parton interactions   in QCD  will   require  
 methods that go beyond the local operator expansion,  and 
    likely    involve  fully  unintegrated parton  correlation  
 functions~\cite{rogers08,*tcrogers}.  
Besides the  relevance  for event generators,   this  should also 
 provides a natural 
framework  for the investigation  at the LHC 
of possible new strong-interaction effects 
at very high energies, including parton saturation~\cite{motyka08}.  

This   report  is based   on the analysis~\cite{hs07,*plb06}  of   
nonlocal operator  expansion,   investigating corrections from 
graphs with multiple  gluon exchange.  The   point of view in this study is 
  to  connect  the treatment of  multi-gluon 
  contributions  with formulations in 
  terms of standard 
partonic  operators, and in  this respect it can be seen as  
 deriving from the approach of~\cite{bukhvostov}. 
We present an illustration for the case  
of structure  functions. This case is   also  treated   
in the analyses of~\cite{bartels99,*Bartels:1999xt,*bartelspeters}. 
More  discussion may be found  in~\cite{npiap}. 
The formulation  discussed below  trades  parton 
distribution functions 
for   moments of eikonal-line correlators.   We expect this  formulation     
  to be  
 useful    also  for  the treatment of  the associated final-state distributions.

\section{From parton distribution functions  
to eikonal-line  matrix elements}
\label{secfrom}

The analysis~\cite{hs07,*plb06} 
starts   with 
the quark distribution function, defined as  
\begin{equation}
\label{eq:fqdef}
 f_{q}(x,\mu) = 
{ 1 \over 4 \pi} 
\int\! d y^- e^{i xP^+y^-}
   \langle P |
\bar\psi(0) Q(0) \gamma^+ 
Q^\dagger(y^-)\psi(0,y^-,{\bm 0})
|P\rangle_c  
\end{equation}
where $\psi$ is the quark field, 
 $Q$ is the gauge link, and 
 the subscript $c$ is the instruction to take 
connected graphs.  
The matrix element (\ref{eq:fqdef}) 
can be rewritten  as the real 
part of a forward scattering amplitude~\cite{hs07}, in which 
we  think of the operator $Q^\dagger \psi$  
as creating an antiquark 
plus an eikonal line  in the minus direction, starting  
at distance $y^-$ from the position of the target. 

Next,  supposing  that $x$ is small,   we  treat  the evolution of the 
antiquark-eikonal system in a  hamiltonian framework (see~\cite{hs07,*plb06} and 
references therein) which 
 allows us to express the evolution operator 
 in the high-energy approximation 
as an expansion in Wilson-line matrix elements. 
The leading term of  this   is  (``dipole"  term)  
\begin{equation}
\label{xidef}
\Xi( {\bm  z}, {\bm b}) = 
\int [ d P^\prime ] \   
\langle P'|\frac{1}{N_c} \ {} {\rm Tr}  \{ 1
 - F^{\dagger}( {\bm b} + {\bm z}/2)\,
F({\bm b}-{\bm z}/2)
\} |P \rangle \hspace*{0.2 cm} ,  
\end{equation}     
where  $F$ is the eikonal operator 
\begin{equation}
\label{Feikon}
F({\bm r }) = {\cal P}\exp\left\{
-ig\int_{-\infty}^{+\infty}dz^- {\cal A}^+_a(0,z^-,{\bm r})   \   t_a
\right\}  , 
\end{equation}
${\bm z}$ is the 
transverse separation between the eikonals in (\ref{xidef}), 
and ${\bm b}$ is the impact parameter. 
 
In this  
  representation   the quark distribution (\ref{eq:fqdef}) 
is given by  the coordinate-space convolution 
\begin{equation}
\label{convfq}
x f_q ( x , \mu) = \int  {d{\bm b}} \ {d{\bm z}} \ 
u (\mu ,  {\bm z} )   \   \Xi( {\bm z}, {\bm b})  
- UV \;\; . 
\end{equation}
 In~\cite{hs07}   the explicit result  is given for the 
function $u (\mu ,  {\bm  z} )$ at one loop  
in dimensional regularization and for the counterterm $- UV$ of   
  $\overline {\rm MS}$  renormalization. 
The $\overline {\rm MS}$ 
result  can  also be recast in a  
physically more transparent form 
in terms of a cut-off on the ${\bm z}$ 
integration region,   as long as 
the scale $\mu$ is sufficiently large compared to the   
inverse hadron radius: 
\begin{equation}
\label{eq:renormalizedfq}
 xf_{q}(x,\mu) 
=   
\frac{N_c}{3 \pi^4}\
\!\int\! {d{\bm b}} \ { {d{\bm z}} \over  { {\bm z}^4 }  }  \  
\theta({\bm z}^2\mu^2 > a^2) \   
\Xi({\bm z} , {\bm b} )  \;\;\; , 
\end{equation}
where $a$ is a renormalization scheme dependent 
coefficient given in~\cite{hs07}.    

 The Wilson-line  matrix element 
 $\Xi({\bm z}, {\bm b})$ 
receives contribution from both long  distances and 
short distances. 
At small ${\bm z}$ it 
  may  be treated by a short distance 
expansion.  At large   ${\bm z}$ it  should   be 
parameterized 
consistently with bounds from unitarity and 
saturation~\cite{motyka08} 
and determined from data.

\section{An    algebraic relation   for     eikonal operators }
\label{secrela}

A general relation 
 between  fundamental and adjoint representation for $\Xi$, valid for any 
 distance  ${\bm z}$, is given in~\cite{hs07}, based on the  algebraic 
 relation 
\begin{eqnarray}
\label{oneminus}
{1 \over {N_c^2 - 1}} {\mbox{Tr}} 
\left[ 1 -  U^\dagger ({\bm z}) U ({\bm 0}) \right] &= &
{ C_A \over C_F}  
 {1 \over {N_c}} \ {\mbox{Re}} \  {\mbox{Tr}} 
\left[ 1 -  V^\dagger  ({\bm z}) V({\bm 0}) \right] 
\\
&-& 
{1 \over 2} { C_A \over C_F}  
 {1 \over {N_c^2}} \ | {\mbox{Tr}} 
\left[ 1 -  V^\dagger  ({\bm z}) V({\bm 0}) \right] |^2 
\hspace*{0.2 cm}   
\nonumber    
\end{eqnarray}
with 
$ V  = F_{\rm{fund.}} $, 
$U  = F_{\rm{adj.}} $.  

From this one can obtain small-$\bm z$ relations  connecting 
$\Xi$ to the gluon distribution. For instance,  
for the fundamental representation  at small $\bm z$  this  yields 
\begin{equation}    
\label{eq:XiofBsmallDelta}
\Xi(\bm b, \bm  z) =  {\bm  z}^2  \    \frac{\pi^2 \alpha_s}{2N_c}\,
x G(x,\mu )\,\phi(\bm b)    ,
\end{equation}
where by $x G$  we denote  the gluon distribution (either   
the $x_c$-scale or weighted-average expressions in~\cite{hs07}),    
and  $\phi(\bm b)$ obeys 
\begin{equation}
\label{b-integral}
\int d\bm b \ \phi(\bm b) = 1   . 
\end{equation}
 The result for $\Xi$ in the fundamental representation corresponds directly 
 to the one for the dipole cross section in the saturation 
 model~\cite{motyka08}. 
 Results in  the fundamental and adjoint cases are relevant  to discuss 
 quark saturation and gluon saturation.

\section{Power-suppressed contributions}
\label{sec:strufun}

In the  s-channel framework of~\cite{hs07} 
contributions to hard  processes  
suppressed by powers of the hard scale 
are controlled by 
 moments 
 of   $\Xi$,  
\begin{equation}  
\label{Mmoments}
  {\cal M} _{p} = {{  2^{2 p} \ p}  \over \Gamma ( 1 - p  ) }
\int   {{d{\bm z}} \over {\pi {\bm z}^2}}
\ ({\bm z}^2 )^{ - p } \
 \int d {\bm b} \ \Xi( {\bm z}, {\bm b}) \; \; ,  
\end{equation}
 analytically continued for  $ p  >  1$.   
Models for   the dipole scattering function   including saturation 
are reviewed in~\cite{motyka08}.    In this case  the moments  (\ref{Mmoments}) are 
proportional to integrals over impact parameter of powers of the saturation 
scale. Higher moments are obtained from derivatives with respect to $p$, 
\begin{equation}
\label{Mmoments1}
  {\cal M} _{p , 0} \simeq  
    \int\! d{\bm b} \ [ Q_s^2({\bm b}) ]^p 
\; \; , 
\;\;\; \; 
  {\cal M} _{p , k} \simeq  (- 1)^k  \  { d^k \over { d p^k}}   {\cal M} _{p , 0} 
\; \;  . 
\end{equation}

As an illustration, 
 we  determine  the    $C_A / x$ part of 
 the coefficients  of  the first subleading power correction 
from the s-channel 
for  transverse and longitudinal 
 structure functions  $F_T$, $F_ L$.    
 Denoting the $Q^2$ derivative by $ \dot F_j = d F_j / d \ln Q^2$ for $ j = T , L$,  
 and its leading-power contribution by  $ {\dot F}_{ j , lead.} $, 
 one has 
\begin{equation} 
\label{bcoeff}
   {\dot F}_{ j }  -    {\dot F}_{ j , lead.}  
=     b_{ j , 0} \   {\cal M} _{2 , 0}   / Q^2   +      b_{ j , 1} \   {\cal M} _{2 , 1}  /Q^2 
 + \cdots 
\;\;\; 
\end{equation}

Structure functions can be analyzed in the same way~\cite{hs07} as 
described  in Sec.~\ref{secfrom} for the quark distribution function. 
The main difference 
 compared to the case of the quark distribution (\ref{eq:fqdef})    is that   
the ultraviolet region of small $\bm z$ is  now  regulated by the physical  scale 
$  Q^2$ rather than requiring, e.g.,    $\overline {\rm MS}$  renormalization. 
Saturation   is  reobtained~\cite{hs07} within  the dipole approximation~\cite{motyka08}. 
By  the analysis based on   (\ref{oneminus}),(\ref{eq:XiofBsmallDelta})  
the saturation   scale   $Q_s(\bm b)$ for a dipole in the 
fundamental representation  is 
\begin{equation}
\label{eq:Qssq}
Q_s^2 ( {\bm b} ) = 
\frac{2\pi^2 \alpha_s}{N_c} \,
x G(x,\mu)\,\phi(\bm b)   .
\end{equation}
To study the expansion  in powers of $1 / Q^2$ it is  
  convenient  to go to  Mellin moment space
by representing  $\Xi$   via the Mellin  transform 
\begin{equation} 
\Xi  ( {\bm z}    ,   \bm b)  = 
 {\bm z}^2 \int_{a - i \infty}^{a + i \infty}  
\frac{du}{2 \pi i}  \ ( {\bm z}^2  )^{- u} \ 
{\widetilde \Xi} ( u, {\bf b} ) 
\;\; , 
\label{Xitilde}
\end{equation}
$0 < a < 1$. 
Then the structure functions $F_{ T , L}$  have the representation 
\begin{equation} 
x   F_{ T , L} 
=    \int  {d{\bm b}}   \int_{a - i \infty}^{a + i \infty}  
\frac{du}{2 \pi i}  \     {\widetilde \Xi} ( u, {\bf b} )   
\   \Phi_{T , L}  (u) 
\;\; , 
\label{Finvmellin}
\end{equation}
where $ \Phi_{T , L}  (u)$  
can  be read  from~\cite{ch94,*cch,*cch90}   and are given by 
\begin{equation} 
    \Phi_T  (u) =    \langle e_a^2  \rangle 
    \frac{N_c }{ 4^{u+2}  \pi^{2} } 
( Q^2  )^{u}   \frac{  \Gamma(3-u) \Gamma(2-u) 
\Gamma(1-u) }{   \Gamma(5/2-u) \Gamma(3/2+u)}  \ (1 + u) \ \Gamma ( u) , 
\label{phiT}
\end{equation}
\begin{equation} 
   \Phi_L  (u) =    \langle e_a^2  \rangle  \frac{N_c }{ 4^{u+2}  \pi^{2} } 
( Q^2  )^{u}   \frac{   [  \Gamma(2-u) ]^3 
 }{   \Gamma(5/2-u) \Gamma(3/2+u)}  \   2  \   \Gamma (1+ u)  , 
\label{phiL}
\end{equation} 
with $\Gamma$ the Euler gamma function. 
The expansion in $1 / Q^2$ of  (\ref{Finvmellin})  is controlled 
  by  the singularity structure   of the integrand 
  in the $u$-plane~\cite{hs07,bartels99,npiap}.  
Eqs.~(\ref{phiT}),(\ref{phiL}) show that 
longitudinal $\Phi_L$ 
has no pole at  $u =0$, so that  the  leading  
  singularity is   given by the $u =0$ pole in ${\widetilde \Xi}$, 
while 
  the first subleading pole $u = -1$    is absent  in transverse 
 $\Phi_T$  due to the numerator factor $(1 + u)$,  so that  
the answer for the transverse case at next-to-leading level  is 
determined by the singularity in  ${\widetilde \Xi}$, with $\Phi$ 
   contributing to  the coefficient  of the residue.   

It can  be verified   that  contributions  to (\ref{Finvmellin}) in 
the lowest 
$ p = 1$  moments   in Eq.~(\ref{Mmoments1})  correctly reproduce  the 
  small-x gluon  part of  renormalization-group  evolution,   
\begin{eqnarray} 
\label{leadpow2}
 {\dot F}_{ T , lead.}   & =  & 
 \langle e_a^2  \rangle 
  \ \frac{\alpha_s}{ 2 \pi  } \int_x^1\!   {dz \over z} \   
 {{  [z^2 + \left(1-z\right)^2 ] }  \over 2 } \,
f_{g}\!\left(\frac{x}{z},Q \right)  + {\rm{quark}} \;\; {\rm{term}}  
\nonumber\\
& \simeq  &   \langle e_a^2  \rangle 
 \  \frac{\alpha_s}{ 2 \pi } \  \frac{1}{ 3 } \  G   
  + {\rm{quark}} \;\; {\rm{term}} 
  \;\; , 
\end{eqnarray}
 using  (\ref{b-integral}),(\ref{eq:Qssq})  and  the  gluon 
 distribution  $G$  evaluated at the average~\cite{hs07}   $x \simeq x_c$, with 
 the lowest $x$-moment  of the gluon $\to$ quark splitting function  
\begin{equation} 
\label{lowmom}
\int_0^1  dz  \ P_{q g} (z) =   \int_0^1  dz \  [ z^2 + (1 -z)^2 ] / 2 = 
1 / 3    \;\;  . 
\end{equation}  
 Beyond leading power,   the first subleading   corrections read  
\begin{equation} 
   {\dot F}_{ T }  -    {\dot F}_{ T , lead.}  
=     - \langle e_a^2  \rangle    \frac{C_A }{ 20 \pi^{3}  x } 
 \frac{1}{ Q^2 } \int\! d{\bm b} \ [ Q_s^2({\bm b}) ]^2   +  \dots  \;\;  , 
\label{firstsubT}
\end{equation}
\begin{eqnarray} 
   {\dot F}_{ L }  -    {\dot F}_{ L , lead.}  
&=&   -  \langle e_a^2  \rangle     \frac{C_A }{ 15 \pi^{3}  x } [  {{14}  \over  {15} }  + \psi (1) ] 
 \frac{1}{ Q^2 } \int\! d{\bm b} \ [ Q_s^2({\bm b}) ]^2   
\nonumber\\
&+&      
 \langle e_a^2  \rangle    \frac{C_A }{ 15 \pi^{3}  x }  
  \frac{1}{ Q^2 } \int\! d{\bm b} \ [ Q_s^2({\bm b}) ]^2 
 \ln  [ { Q^2  /    {Q_s^2({\bm b})   }   } ]   
 + \dots  \;\;   .  
\label{firstsubL}
\end{eqnarray} 
That is,  
the $b$ coefficients in (\ref{bcoeff})  are given by 
\begin{eqnarray} 
\label{bcoeffres}
b_{ T , 0} =  - \langle e_a^2  \rangle  \   C_A /  (20 \pi^{3}  x)    \;  & , &  \;\;
b_{ T , 1} =  0     \;\;   ,   
\nonumber\\ 
b_{ L , 0} =  - \langle e_a^2  \rangle  \   C_A  \ 
[14/225 + \psi(1) / 15 ] /  ( \pi^{3}  x)    \; & , &  \;\;
b_{ L , 1} =      \langle e_a^2  \rangle  \  C_A /  (15 \pi^{3}  x) 
 \;\;   ,   
\end{eqnarray} 
 with $\psi$ the Euler psi function.

Via process-dependent coefficients analogous  to those in (\ref{bcoeff}),   the   
eikonal-operator moments   (\ref{Mmoments}) will also control power-like  
contributions to  the associated jet cross sections 
   due to multi-parton interactions 
in the initial state. 
At present these processes are modeled by  Monte Carlo, which   point  to their  
quantitative 
significance    for   the proper  simulation of  hard events at the LHC. 
The  above discussion 
 suggests the potential usefulness  in this context  
 of analyzing jet and structure 
function data  by trading 
 parton distribution functions for  s-channel correlators defined according to 
 the  method of Sec.~\ref{secfrom}.

\bibliographystyle{heralhc} 
{\raggedright
\bibliography{heralhc}
}
\end{document}